\input{epsf}
\documentstyle[11pt,paspconf]{article}

\def\kms{\rm ~km~s^{-1}}
\def\etal{{\it et al. }}

\def\etal{{\it et al. }}
\def\gsim{ \lower .75ex \hbox{$\sim$} \llap{\raise .27ex \hbox{$>$}} }
\def\lsim{ \lower .75ex \hbox{$\sim$} \llap{\raise .27ex \hbox{$<$}} }
\def\pp{\noindent\parshape 2 0truecm 12.0truecm 1.5truecm 10truecm}

\begin{document}

\title{The Structure of Cold Dark Matter Halos}

\author {Ben Moore, Sebastiano Ghigna, Fabio Governato}

\affil{Department of Physics, Durham University, South
Road, Durham City, DH1 3LE, UK} 

\author{ George Lake, Tom Quinn \& Joachim Stadel }

\affil{Department of Astronomy, University of Washington,
Seattle, WA, USA}

\begin{abstract}
We investigate the internal structure of cold dark matter halos 
using high resolution N-body simulations. As the mass and force 
resolution are increased, halo density profiles become steeper, 
asymptoting to a slope of $\sim -4/3$ in the central regions 
and may not have converged to a unique result.
At our highest resolution we have nearly 3 million
particles within the virial radius, $R_{200}$, and force softening
that is $\sim 0.2\% R_{200}$.  
This resolution has also allowed us to
resolve a large part of the overmerging problem \-- we find over 1000
surviving dark halos orbiting within a single cluster potential.  These
data have given us unprecedented insights into the dynamics and
structure of ``halos within halos'', allowing us for the first time,
to compare the distribution of dark matter with observations of
galaxies in clusters.
\end{abstract}


\section{Introduction}

We follow the path of many previous researchers who have used
numerical simulations to investigate the formation and structure
of dark matter halos ({\it c.f.}  Zurek
\etal (1986), Frenk \etal (1988), Dubinski \& Carlberg (1991), Warren
\etal (1992), Carlberg (1994), Crone \etal (1994), Summers \etal (1995) 
Navarro, Frenk \& White (1996, NFW), Tormen \etal (1996). 
The systematic study by NFW showed that
halos density profiles may follow a universal form, uniquely
determined by their mass and virial radius; varying from $r^{-1}$ in
the central regions, smoothly rolling over to $r^{-3}$ at the virial
radii. These studies have a resolution such that halos typically
contained $\sim 10,000$ particles and force softening of $1\%$ of the
virial radii. (The higher resolution simulation by Dubinski \&
Carlberg (1991) showed the same steep inner profiles, but
had open boundary conditions such that infall ceased at $z\sim 1$. )

The CDM model has power on all scales, allowing most of the mass to
collapse into very small halos at early epochs that are not resolved
by numerical simulations. If this material at higher densities can be
resolved, then it may alter the evolutionary history of the final
halo, perhaps leading to different density profiles.  One of the key
aims of this work is to verify the convergence to a unique profile.

Moore, Katz \& Lake (1996) pointed out that the amount of substructure
that could survive within virialised regions depended simply upon
the numerical resolution and that the overmerging problem resulted
primarily from large softening lengths. Halos in high density
environments would be tidally disrupted since they are artificially
``soft''.  One of the results of the higher mass and force resolution
is that halos can be found surviving within high density regions and
thus for the first time we can examine the dynamics and properties of
the dark halos within clusters.

\section{The simulations}

A candidate cluster is initially identified from a large cosmological
volume that has been simulated at lower resolution. The particles
within the selected halo are traced back to the initial conditions to
identify the region that will be re-simulated at higher
resolution. The power spectrum is extrapolated down to smaller scales,
matched at the boundaries such that both the power and waves of the
new density field are identical in the region of overlap, then this
region is populated with a new subset of less massive
particles. Beyond the high resolution region the mass resolution is
decreased in a series of shells such that the external tidal field is
modeled correctly in a cosmological context. The new initial
conditions are integrated with high accuracy using the parallel
treecode PKDGRAV, using variable timesteps and periodic boundary
conditions. We systematically increase the force and mass resolution
for two halos with virial radii of 2 Mpc and 3.4 Mpc. At our highest
resolution our force softening is 5 kpc and the particle mass is
$8\times 10^8M_\odot$ within a cluster with $R_{200}=3.4$ Mpc and mass
$\sim 2\times 10^{15}M_\odot$ (we assume H=50$\kms$ Mpc$^{-1}$ throughout).

\begin{figure}
\begin{picture}(350,350)
\put(20,130)
{\epsfxsize=8.0truecm \epsfysize=8.0truecm 
\epsfbox[0 0 600 700]{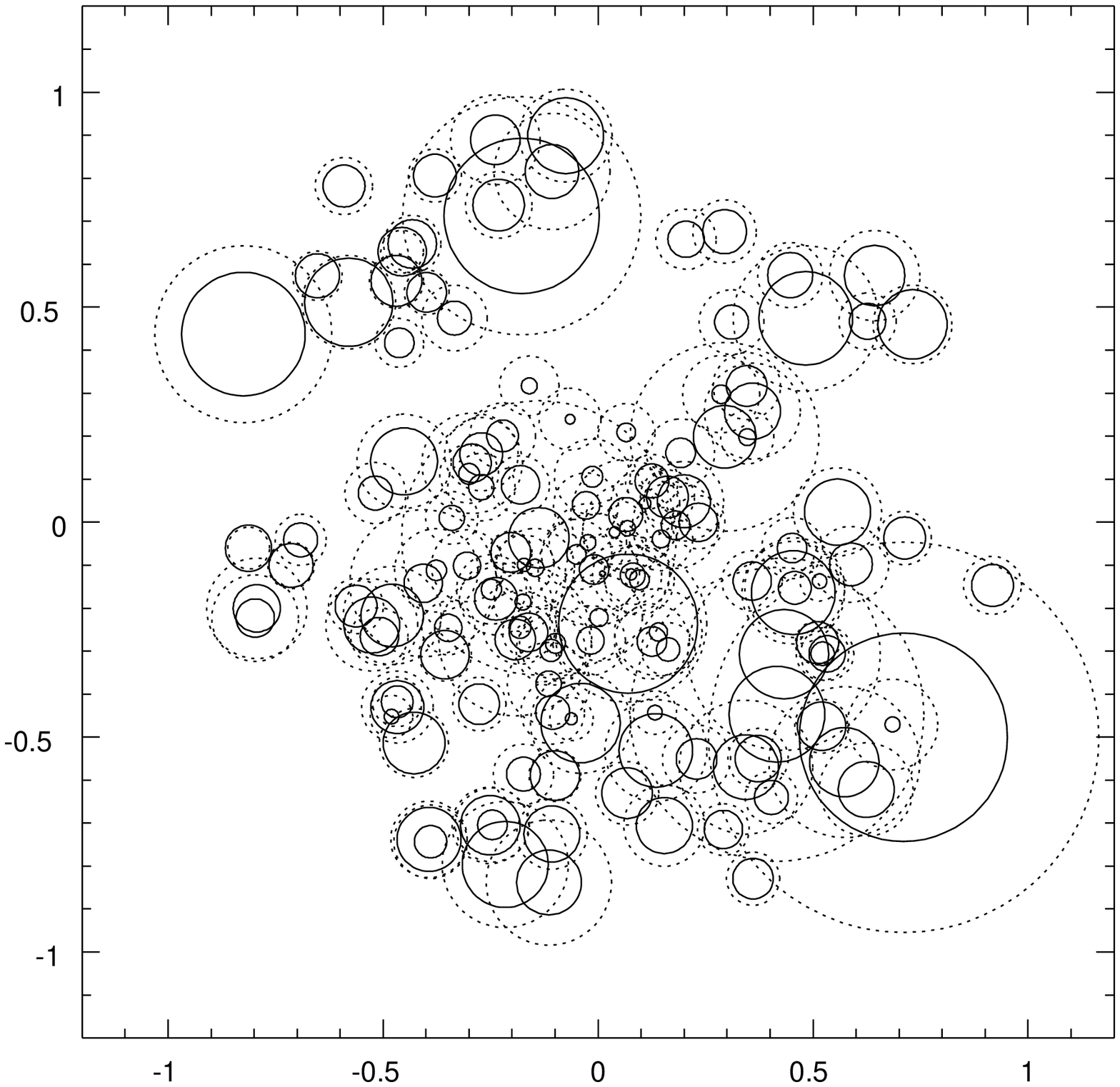}}
\put(20,-50)
{\epsfxsize=8.0truecm \epsfysize=8.0truecm 
\epsfbox[0 0 600 700]{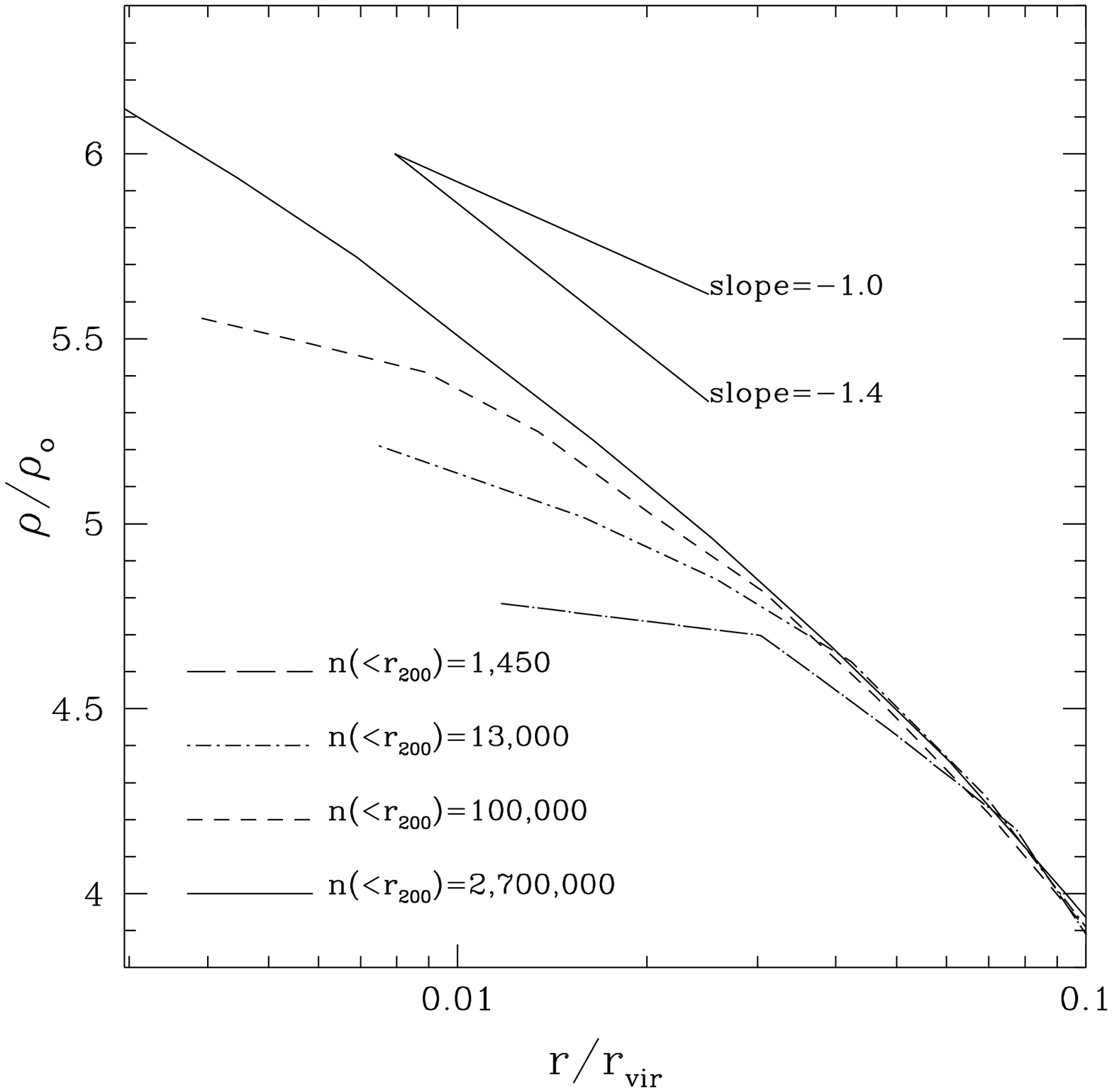}}
\end{picture}

\caption[junk]{ 
The upper panel shows the
projected distribution of substructure halos that lie within $R_{200}$.
The solid circles give the measured tidal radii of the halos
and the dotted circles denote their theoretical virial radii
calculated assuming they are isothermal spheres. Mass
loss via tidal stripping and halo-halo collisions is
clearly evident. The length unit in this plot is in units
of the cluster virial radius. Note that the covering
factor of dark matter substructure is of order unity.
The lower panel shows the inner density profiles of the entire 
cluster halo as we increase the force and mass resolution. 
The curves start at the values of the softening parameter used.
As a rule of thumb, we use spline softening lengths that
are 1/50th of the mean inter-particle separation within the whole box.
Comparing the halo profiles at different resolutions demonstrates that 
we can ``believe'' a given profile to a scale of about 0.5 times 
the mean inter-particle separation of those particles within the 
virial radius. Note that this is a smaller scale than advocated by
Splinter \etal (1997) who argue that many statistics are affected on
scales smaller that the global mean interparticle seperation.

}
\label{slice}
\end{figure}

\section{Results}

\noindent

As we increase the mass and force resolution we find that the inner
slope of the density profiles become as steep as -1.4.  
We suspect that the steepening of the slope from -1.0
found at lower resolution, results because we are resolving
many more halos collapsing at early epochs when the mean density 
of the Universe was high. This material is more robust against 
tidal disruption and can sink deeper within the cluster potential.
Alternatively, Evans \& Collett (1997) demonstrated that an inner
slope of -4/3 is a stable solution to the Fokker-Planck equation
and collisional Boltzman equation.

Halos on galaxy scales will collapse earlier than the cluster
simulated here, thus at a fixed radius measured in terms of
the virial radius, galaxy halos will be even denser than clusters.
These steep inner profiles will lead to problems when
trying to reproduce the properties of giant arcs in clusters,
the slow rise of stellar velocity dispersions with radius in Cd
galaxies and the rotation curves of disk dominated galaxies.

The distribution of halos orbits are close to
isotropic; the median value of apocentric to pericentric distances is
5:1, a value that does not vary with position within the cluster and
is unbiased with respect to the orbits of the smooth particle
background. Circular orbits are rare and more than 25\% of all halos
within the cluster will pass within 200 kpc $\equiv 0.1R_{200}$.

Dark halos will be tidally truncated to a value determined by
the density of the cluster at their pericentric positions.  The
approximation of isothermal halo mass distributions orbiting within a
deeper isothermal potential works very well; {\it i.e.}
$r_{tidal}=r_{peri}\sigma_{halos}/\sigma_{clus}$.  

The mass bound to resolved dark matter halos is approximately 10\% of the
entire cluster mass and varies from 0\% near the cluster center, to 20\%
at its virial radius. This latter value is roughly the expected value
for the mass attached to halos above a circular velocity of 80 $\kms$.
(As numerical resolution increases, we might expect this number to
asymptote to unity.)  Correspondingly, through tidal stripping, the
sizes of halos vary with cluster-centric radius, an effect that will
be readily observable using weak lensing of background galaxies.

Overmerging within the central regions of dense halos
leads to a final distribution of substructure that is always 
anti-biased with respect to the global mass distribution. 
We find no evidence for a velocity bias \-- halos move
on the same orbits as the smooth particle background and
have the same velocity dispersion.


The density profiles of a sample of well resolved 
halos within the cluster and those in
the cluster proximity,  have significantly higher concentrations than
those found in isolated environments. This is 
due to their earlier collapse redshifts rather than the
internal response of the halos to mass loss and heating from tidal
stripping. 
 
Most of the halos within the cluster and in the cluster proximity have
density profiles that are well fit by NFW profiles.  Halos that loose
a great deal of mass through tidal stripping have outer density
profiles as steep as $\rho(r) \propto r^{-4}$ (at $\approx 30\%$ of
their virial radius), thus Hernquist profiles provide slightly better
fits.

Mergers between halos with mass ratios larger than 10:1, occur with a
frequency of about 20\% in the cluster vicinity over the past 5 Gyrs.
Once within the virial radius mergers are very rare; not a single
merger occurs for halos within 90\% of $r_{200}$ of the cluster center.


\begin{references}

\pp Carlberg R. 1994, {\it Ap.J.}, {\bf 433}, 468.

\pp Crone M.M., Evrard A.E. \& Richstone D.O. 1994, {\it Ap.J.}, {\bf 434}, 402.

\pp Dubinski J. \& Carlberg R. 1991, {\it Ap.J.}, {\bf 378}, 496.

\pp Evans W.N. \& Collett J.L. 1997, {\it Ap.J.Lett.}, in press.

\pp Frenk C.S., White S.D.M., Davis M. \& Efstathiou G. 1988, {\it Ap.J.}, 
{\bf 327}. 507.

\pp Moore B., Katz N. \& Lake G. 1996, {\it Ap.J.}, {\bf 457}, 455.

\pp Navarro J.F., Frenk C.S. \& White S.D.M. 1996, {\it Ap.J.}, {\bf 462}, 563.

\pp Quinn P.J., Salmon J.K. \& Zurek W.H. 1986, {\it Nature}, {\bf 322}, 329.

\pp Splinter R.J., Melott A.L. and Shandarin S.F. 1998, {\it Ap.J.}, in press.

\pp Summers F.J., Davis M., \& Evrard A.E. 1995, {\it Ap.J.}, {\bf 454}, 1.

\pp Tormen G., Bouchet F.R. and White S.D.M. 1996, {\it M.N.R.A.S.}, 
{\bf 286}, 865.

\pp Warren S.W., Quinn P.J., Salmon J.K. \& Zurek H.W. 1992, {\it Ap.J.}, 
{\bf 399}, 405.

\end{references}
\end{document}